\documentclass[]{aa}

\usepackage{graphicx}
\usepackage{txfonts}
\usepackage{natbib}
\bibpunct{(}{)}{;}{a}{}{,}

\def\pcite#1{[ref]}

\begin{document}
   \title{Cluster Geometry $\&$ Inclinations from Deprojection Uncertainties}
   \subtitle{Cluster Geometry $\&$ Inclination}
   \author{Dalia Chakrabarty \inst{1},
          Elisabetta De Filippis \inst{2}
          \and Helen Russell \inst{3}
          }
   \offprints{Dalia Chakrabarty}
   \institute{School of Physics $\&$ Astronomy,
              University of Nottingham, 
              Nottingham NG7 2RD, U.K.
              \email{dalia.chakrabarty$@$nottingham.ac.uk},
	      Department of Physical Sciences, 
	      University of Napoli Federico II,
	      via Cintia 6, 80126 Napoli, ITALY
              \email{betty@na.infn.it}
              \and
	      Institute of Astronomy,
	      University of Cambridge,
	      Madingley Road,
	      Cambridge CB3 0HA.
              \email{hrr27@ast.cam.ac.uk}
             }
\authorrunning{Chakrabarty, {De} Filippis $\&$ Russell}
\titlerunning{Cluster Geometry $\&$ Inclination}

   \date{\today}

\abstract {The determination of cluster masses is a complex problem
that would be aided by information about the cluster shape and
orientation (along the line-of-sight).} {It is in this context, that we
have developed a scheme for identifying the intrinsic morphology and
inclination of a cluster, by looking for the signature of the true
cluster characteristics in the inter-comparison of the different
deprojected emissivity profiles (that all project to the same X-ray
brightness distribution) and by using SZe data when available.} {We
deproject the cluster X-ray surface brightness profile under the assumptions
four different ellipsoidal geometry and inclination configurations
that correspond to four extreme scenarios, by the non-parametric
algorithm DOPING. The formalism is tested with model clusters and then
is applied to a sample of 24 clusters. While the shape determination
is possible by implementing the X-ray brightness alone, the estimation
of the inclination with the line-of-sight is usually markedly improved
upon by the usage of SZe data that is available for the considered
sample.}{We spot 8 prolate systems, 1 oblate and 15 of the clusters in
our sample as triaxial. In fact, for systems identified as triaxial,
we are able to discern how the three axis ratios compare with each
other. This, when compounded by the information about the
line-of-sight extent, allows us to constrain the inclination quite
tightly.}{}


\maketitle

\section{Introduction}
\noindent
The identification of the three dimensional cluster shape is of
significant importance in the pursuit of quantities of cosmological
interest. In particular, the misinterpretation of the true cluster
geometry can affect the extraction of the Hubble constant from X-ray
and SZe information \citep{zaroubi01} and of the cluster
mass. Erroneous cluster masses are often the cause of less reliable
understanding of cluster physics and poorer constraints from cluster
cosmology. A mandatory exercise that should be undertaken towards the
extraction of the correct cluster masses is the identification of the
correct geometry and inclination of the cluster. However, the lack of
knowledge about these attributes are typically bypassed by resorting
to the assumption of sphericity, even when the projected ellipticity
of the system is indicated to be non-zero from observations
\citep{fabian81, yoshikawa99, pizzolato03}. This is also the adopted
geometry in frequently used packages for determining dark matter
distributions in clusters, like JACO \citep{jaco}.

On the contrary, halos realised in cosmological simulations have been
found to be flattened and triaxial in shape; the distribution of
ellipsoidal shapes of simulated halos have been discussed by
\cite{frenk_88, dubinski_91, warren_92, cole_lacey, bailin_steinmetz},
among others. Moreover, the formation of large scale structure has a
bearing on the intrinsic shape of cluster sized halos
\citep{plionis04}; thus, the 3D cluster morphology could be a device
that can be used to constrain cosmological models. 

While the statistical analysis of the 3D cluster morphology has been
looked into \citep[see][and references
therein]{jing02,kasun05,hayashi07}, the determination of the full 3-D
shapes and orientations of individual halos is a harder problem that
has attracted relatively less attention \citep{sereno06}.

Recovery of the full 3-D morphology of a cluster, using measurements of
accessible cluster characteristics (such as X-ray emission profiles or
2-D X-ray surface brightness distributions or SZe data), would require
expertise in deprojecting such observed information into the full 3-D
distribution, under general triaxial geometries. Such can be achieved
via parametric fits; in fact, parametric deprojection of the observed
optical surface brightness maps of galaxies, under assumptions of
sphericity and (more rarely) axisymmetry, has been studied before
\citep{palmer, bendinelli}. However, the fundamental problems with
parametric fits are (i) the answer depends on the choice of the
parametrisation and (ii) the goodness-of-fit quantifier (such as a
$\chi^2$ measure) can appear spuriously inflated, particularly in the
presence of non-homogeneous measurement noise \citep{bissantz_munk}.

Thus, non-parametric deprojection is a better option. However,
deprojection algorithms that promise improved three dimensional
cluster mass distributions, by taking the measured ellipticity and the
true cluster morphology and orientation into account, are limited in
availability. \cite{zaroubi01} report on the application of a
non-parametric deprojection algorithm that assumes axisymmetry, to a
set of simulated clusters. However, as the authors state, the
applicability of this scheme to the current state of data sets appears
difficult.

A comprehensive approach that calls for the amalgamation of two or
more mass indicators include the implementation of X-ray observations,
SZe information, lensing results and dynamical measurements. Such
exercises have been undertaken already
\citep{zaroubi98,reblinsky00,sand02}.  An attempt has been made in the
recent past \citep{sereno06} to decipher the three dimensional
morphology and inclination distribution of a sample of clusters from
\cite{reese02}, with the aim of improving the cluster mass estimation
(from X-ray measurements) more accurately. This work employs the
rudimentary SZe temperature decrement data \citep{betty05} in
the $\beta$-modelling of the cluster. Thus, this work is susceptible
to very large error bars that currently plague SZe data. This is of
course topped by the errors that can be introduced by the choice of
ellipsoids of revolution that are implemented to model their sample
clusters.

In this work, we present a novel, model-independent trick to determine
the correct morphology and inclination of a cluster, without resorting
to the assumption of axisymmetry. Such a determination is made
possible by analysing the {\it multiple} deprojected emissivity or
X-ray luminosity density distributions that are recovered under
distinct assumptions about the cluster geometry and inclination and
that will all project to the same observed X-ray brightness map. Thus,
the deprojection in question needs to be performed by a scheme that is
efficient in carrying out deprojection under general geometries, at
any assumed inclination.  This is possible with a new inverse
deprojection algorithm DOPING (Deprojection of Observed Photometry
using an INverse Gambit) that has been reported in \cite{doping}.

The {\it inter-comparison of the amplitudes and shapes of the
different deprojected emissivity distributions} tell us the correct
shape of the cluster (including triaxial ones) while the inclination
constraints are improved upon by the implementation of the available
cluster elongation information from the SZe data, as reported in the
literature. In case of triaxial systems, the SZe data is used to
determine the two intrinsic axial ratios.

The paper is organised as follows. Following the first introductory
section, we discuss the deprojection algorithm in brief. Section~3 is
devoted to the methodology that we use to extract the cluster shapes from
the deprojected profiles. The testing of the advanced scheme is dealt
within the following section. We then proceed to apply this method to
our cluster sample (Section~5), which precedes the results
section. The paper is rounded off by recounting some relevant aspects
of the presented work.

\section{DOPING}
\label{sec:doping}
\noindent
DOPING stands for Deprojection of Observed Photometry using an INverse
Gambit. However, the name does not do full justice to the capacity of
the algorithm; it is a deprojection algorithm that could just as well
be applied to deproject observed X-ray 2-D surface brightness maps
($SB$) of clusters. The deprojection can be performed in {\it general
triaxial geometries and is able to incorporate radial variations in
shape}. The deprojected luminosity density distribution is sought by
penalised likelihood approach that uses an MCMC optimiser. Although
DOPING is in general able to use the 2-D $SB$ information and provide
the full three dimensional $\rho$, in the case of an oblate/prolate
system with a uniform projected eccentricity, the effective
dimensionality of the problem reduces by one, since the projected and
intrinsic eccentricities can then be related through analytical
relations. This idea is exploited in the work. The $SB$ is treated as
a function of distance along a photometric axis (the ${\bf
x}$-axis). Similarly, the emissivity or X-ray luminosity density
distributions are presented along the ${\bf x}$-axis.

\section{Method}
\noindent
As explained in Section~\ref{sec:doping}, DOPING is able to
incorporate information about the LOS extent of a cluster from SZe
measurements, in the process of deprojecting under the assumption of
triaxiality. However, the poor quality of the currently available SZe
data will cause the determination of shape ambiguous.  Thus, for the
current purpose, we will be carrying out the deprojection of the
observed X-ray $SB$ under assumptions of prolateness or oblateness for
the intrinsic shape of the observed cluster; as for its inclination,
we choose to deproject at the two ends of the range of inclinations
allowed for an oblate geometry (discussed below). The other, even more
important reason for choosing to deproject at these geometry+inclination
configurations is the relative simplicity of deprojection under
oblateness/prolateness than triaxiality.

The different X-ray luminosity density distributions or the
emissivity distributions ($\rho$) that will be recovered from such
deprojections, will be analysed to tell us if the system under
consideration is prolate, oblate or even triaxial.

Now, let us examine our choice of inclinations for deprojection,
in greater detail. We will carry out the deprojections under the
assumptions of $i$=90$^{\circ}$ and $i$=$i_{min}$, where $i_{min}$ is
the smallest allowed inclination for the measured projected
eccentricity for the assumption of oblateness ($i_{min}=\sin^{-1}e_p$,
where $e_p$ is the projected eccentricity that is related to the
projected axial ratio $q_p$ as $e_p^2=1-1/q_p^2$). As a result of such
deprojection of an $SB$, we will be able to say if the cluster is
inclined to the line-of-sight (LOS) by an angle that is less than
$i_{min}$ or is of a value intermediate to 90$^{\circ}$ and $i_{min}$.

Thus, there are 4 distinct deprojection configurations ($D$) that we will
consider for any given SB:
\begin{enumerate}
\item assumed geometry is prolateness and assumed $i$=90$^{\circ}$:
$D^{p}_{90}$,
\item prolateness and $i$=$i_{min}$: $D^p_{min}$, 
\item oblateness and $i$=90$^{\circ}$: $D^o_{90}$ and
\item oblateness and $i$ = $i_{min}$: $D^o_{min}$.
\end{enumerate}
On the basis of these deprojections, we will identify the shape and
inclination of a given cluster.

\subsubsection{Effect of $D$}
\noindent
It is to be emphasised that our determination of the true 3-D shape of
a cluster is {\it not} affected by the exact geometry+inclination
configuration that we choose to deproject its X-ray $SB$ under;
rather, it is the intercomparison of the various profiles of $\rho$
that are recovered from deprojection under the different
geometry+inclination configurations that we choose. Thus, if we choose
to deproject an observed X-ray $SB$, assuming the cluster to have a
geometry+inclination different from the ones that we choose, then the
inter-comparison of the corresponding profiles of $\rho$ will be
different from what we infer here. It is a question of standardising
such an inter-comparison, in order to be able to conclude the true
cluster characteristics. All we do in this paper is advance the same,
using the trends in the inter-comparison that we notice, when the
cluster at hand is deprojected under our choices of
geometry+inclination, i.e. $D$. The modus-operandi that connects such
an inter-comparison to the true cluster characteristics, is discussed
below (Section~\ref{sec:under}).

\subsection{Coordinate System}
\noindent
The two coordinate systems that suggest themselves readily to any
projection related investigation are the body coordinates of the
observed system (considered regular Cartesian, marked here in upper
case letters $X-Y-Z$) and the observer's coordinate system (marked in
lower case $x-y-z$). The POS is spanned by the $x-y$ plane while $z$
is along the LOS. Also, we consider one of the principal axes of the
cluster ($X$) to be coincident with one of the photometric axes $x$
(the photometric major axis for oblate models and the photometric
minor axis for the prolate case).

\subsection{Underlying Principle}
\label{sec:under}
\noindent
The determination of a cluster's intrinsic shape and inclination is
based on the fact that the {\it inter-comparison} of $\rho$ recovered
by deprojecting under the assorted geometry and inclination
configurations $D$, bears the signature of the true characteristics of
the cluster. 

The comparison of (the shape and amplitude of) two distinct profiles
of $\rho$ that will be recovered from deprojecting a given X-ray $SB$,
under two distinct deprojection scenarios (2 different $D$s), will
depend on the inter-comparison of the 
\begin{itemize}
\item extents along the LOS, of the 2 systems that are ascribed to the
2 different assumptions of geometry (corresponding to the 2 $D$s) 
\item ellipsoidal radii to the same generic point ($x,y,z$), in the 2 systems
described by the 2 $D$s, assuming the gas density to be stratified on
concentric ellipsoidal shells.
\end{itemize}
In this section, these ideas are expounded upon.

\subsubsection{Effect of Varying the LOS Extent} 
\label{sec:LOS}
\noindent
The luminosity density that is required to be projected through a
given length, in order to produce an observed $SB$, must be higher
when the extent along the axis of projection is less in a deprojection
model than in the case of the ``true system'', i.e. the observed
cluster.  In other words, $LOS_D < LOS_T\Longrightarrow \rho_D >
\rho_T$. Here, the subscript ``$D$'' refers to quantities relevant to
the deprojection model while the subscript ``$T$'' refers to
quantities relevant to the true system. (We will follow this notation
elsewhere in the paper too).

\subsubsection{Effect of varying the 3-D Density Structure}
\label{sec:main}
\noindent
We can develop a qualitative understanding of the shapes of the
recovered density profiles by assuming that the density is given as a
function of the {\it ellipsoidal radius} $\xi$. The form of $\xi$ is
determined by the geometry and inclination $i$ that describe a system, as
well as its intrinsic eccentricity $e$. Thus, for an oblate system,
inclined at angle $i$, for any $(x,y,z)$, $\xi$ is:
\begin{eqnarray}
\displaystyle{
              x^2 +y^2\displaystyle{(\cos{^2}{i} + \frac{\sin{^2}{i}}{1-e^2})}
              +z^2\displaystyle{(\sin{^2}{i} + \frac{\cos{^2}{i}}{1-e^2})}}
& & \\ \nonumber
\displaystyle{-yz\sin{2i}\displaystyle{\frac{e^2}{1-e^2}}
              } &=& \xi^2.
\label{eqn:ell_i}
\end{eqnarray}
Thus, we assume $\rho=\rho(\xi)$. We also assume that the physically
motivated idea of monotonicity holds true in the density distribution:
$\rho(\xi_i) < \rho(\xi_j), \forall \xi_i > \xi_j$.  At this stage,
this exercise is held independent of effects of LOS extent on the
density.

Now, let us consider two systems with different $e$, $i$ and
geometries. 
\begin{itemize}
\item Let the systems be such that {\it in System~I, the
ellipsoidal radius} to a general point $x,y,z$ is higher than in
System~II, i.e. to $x,y,z$, $\xi_I > \xi_II$ (say)
\item Then $\rho_I (x,y,z) < \rho_II (x,y,z)$, since we assume
density to monotonically fall with $\xi$.
\item Since the projection of density (observed $SB$) is given from
measurement, to compensate for this trend, in System~I, a given value
of density would show up at a higher $\xi$, than in System~II.
\end{itemize}
This translates to the given density showing up at a higher $x$, in
System~I than System~II, when the density profile along the $\hat{\bf
x}$-direction is sought.

Now let us replace System~II with the true system and let System~I be
one of the 4 deprojection models that we consider. Thus, in this case,
$\xi_D > \xi_T$. This comparison can imply stronger effects on the
density at smaller values of $x$ or larger values of $x$, depending on
the relative geometries of the two systems in question. While such
situations are explored below, now let us monitor the consequences of
$\xi_D$ exceeding $\xi_T$, first near the core of the cluster and then
at the outer parts of the cluster.

\begin{figure*}
\centering
\includegraphics[width=13cm]{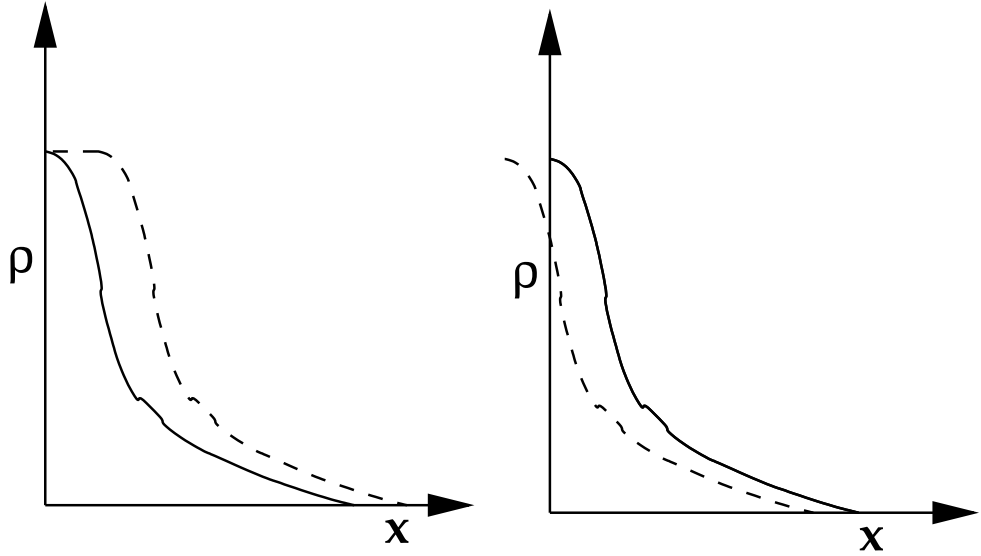}
\caption{\small{If the ellipsoidal radius to most points on the $z=0$
plane in the deprojection model is less than that in the true system,
then density of the deprojection model will in general be higher than
in the true system. In this case, the density profile recovered along
the ${\bf x}$-axis will be characterised by the same value of density
appearing at smaller values of $x$. This results in a deprojected
density profile that is as shown in broken lines on the right
panel. The true density profile of the observed cluster is shown in
the same panel, in solid lines. The deprojected profile is shown
schematically as simply the amplitude-lessened version of the true
profile and therefore is shown to extend to negative $x$-values; in
reality, deprojected density values are recovered only for positive
$x$. On the other hand, if $\xi_D$ exceeds $\xi_T$ on the average,
then deprojected density will fall short of the true density, which
will imply a larger core size in the profile recovered in this case
(shown in broken lines on the left).}}
\label{fig:profiles1}
\end{figure*}

When $\xi_D > \xi_T$ around the core, it has the following effect on
the shape of the recovered density profile. If the density profile of
the true system is as shown in solid lines on the left in
Figure~\ref{fig:profiles1}, then the recovered density profile will be
flatter at the centre, as represented by the broken lines in this
figure. If the smallest $x$ at which density information is given for
the true system is $x_0$, then that for the deprojection model will be
$x_1$, where $x_1 > x_0$, for a given choice of the radial binning. In
lieu of any information available for the radial bins inner to $x_1$,
the code will spread light out uniformly over these bins, using the
same density that is recovered at $x_1$. Thus, the {\it deprojected
density profile will appear to manifest a larger core than the true
density profile}.

\begin{figure}
\centering
\includegraphics[width=8cm]{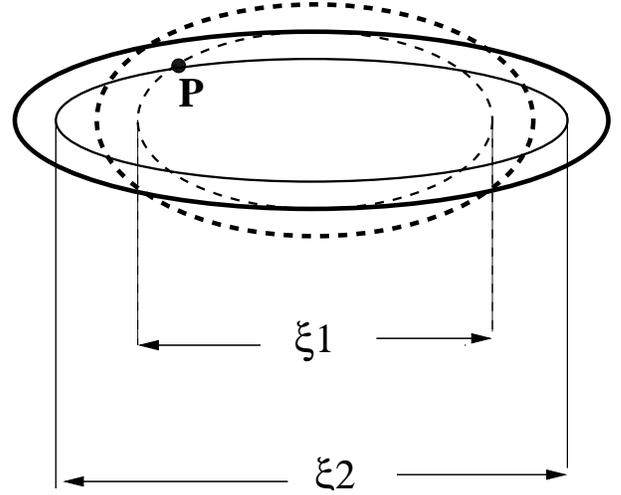}
\caption{\small{The ellipsoidal radius to a point P is marked as
$\xi1$ and $\xi2$, for the two different ellipsoidal configurations,
the projection on the plane of the paper of which are shown in thick
solid and thick broken lines. The system with the bigger long axis and
the smaller short axis corresponds to a higher ellipsoidal
radius. Along with the variations in depth, differences in ellipsoidal
radius is the other factor that is affected by the choice of the
deprojection model.}}
\label{fig:ell_rad}
\end{figure}

However, when the inequality: $\xi_D > \xi_T$ is stronger at larger
than at smaller $x$, the deprojected density profile will end up being
flatter on the outside than the true system.

In contrary to this case, if we have a deprojection model that
generally corresponds to a lower $\xi$ at any $x,y$ on the POS, than
in the true system, the code will attribute a given density to a lower
value of $x$ in the density profile recovered with the deprojection
model than in the true system. The deprojected density profile will
then appear to manifest a similar shape to the true density profile,
but will have a comparatively lower amplitude. This case is shown in
the right panel of Figure~\ref{fig:profiles1}.

In general, we can base our understanding of the density structure in
a cluster, by relying on the following ideas that are illustrated in
Figure~\ref{fig:profiles1} and Figure~\ref{fig:ell_rad}:
\begin{itemize}
\item In two prolate systems with the same long axis, the cluster with
the relatively smaller short axes will imply a bigger $\xi$ at a
general point, as long as the point in question lies close to the long
axis of the system.
\item Similarly, in two prolate systems with the same short axes, the
bigger is the long axis, bigger is $\xi$ to a general point, as long
as this comparison is sought within the extent of the smaller of the
two long axes in question.
\item In two oblate systems with the same long axes, the smaller the
short axis, the bigger is $\xi$ to a general point $x,y,z$, (the
point is considered a part of both systems).
\item Similarly, when we consider two oblate systems with the same
short axes, the cluster with the bigger long axes corresponds to a
bigger $\xi$.
\item In general, it is true that a prolate system with the same axial
ratio as an oblate system, will imply a larger $\xi$ to the point
$x,y,z$ than the oblate system.
\item It is also generally true that deviations of an observed cluster
from a prolate deprojection model will affect the outer parts of the
density profiles while differences between an observed system and an
oblate model implies differences in the recovered density profile that
will manifest themselves at lower values of $x$.
\end{itemize}

\subsubsection{True system: $T^p_i$}
\noindent
Let us elucidate the effects discussed in Section~\ref{sec:LOS} and
Section~\ref{sec:main}, in reference to the true system $T^p_i$,
i.e. a cluster that is prolate in shape and is inclined at a angle $i|
i_{min} < i < 90^\circ$. The projected axial ratio is $q$ and the
extent along the photometric semi-major axis is $a$. 

\begin{figure*}
\hskip-2cm
\includegraphics[width=18cm]{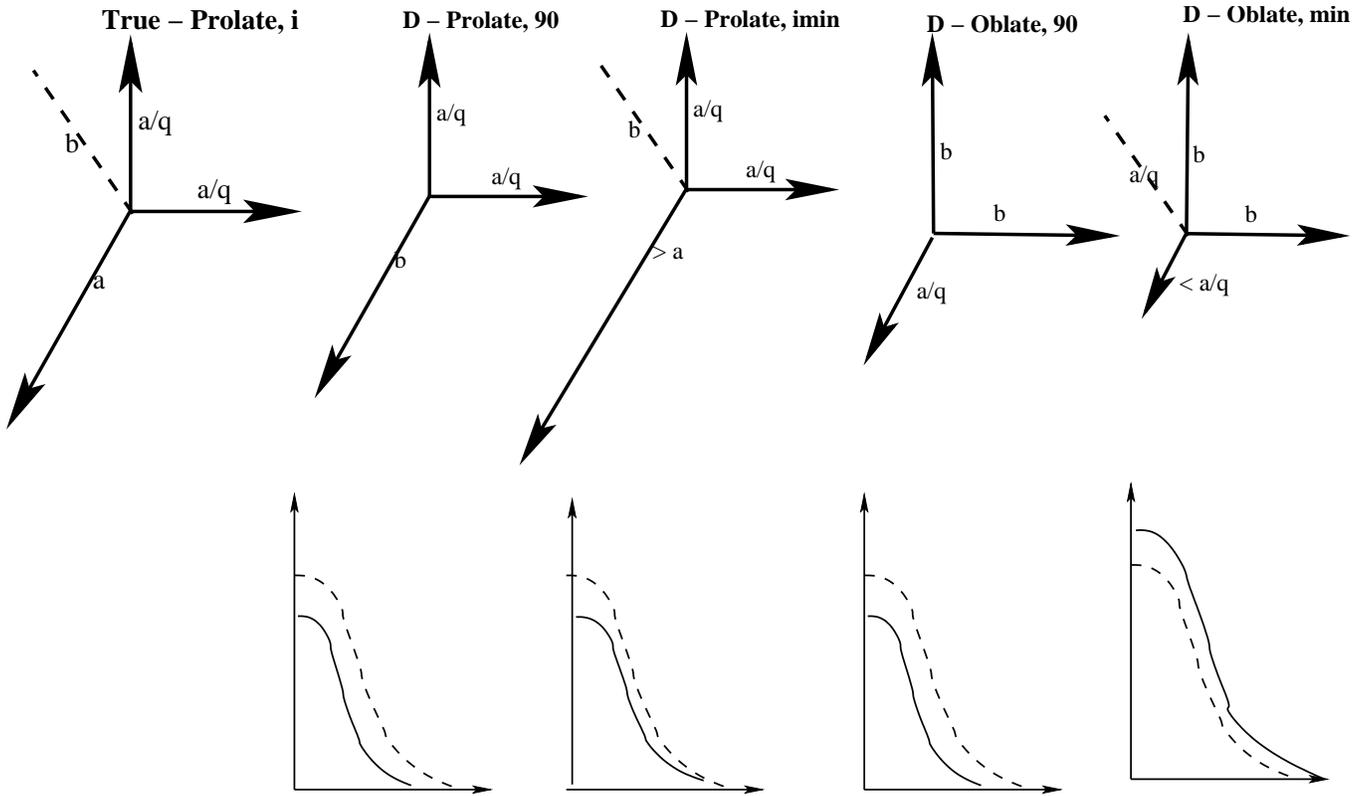}
\caption{\small{Figure showing interpretation of the observed
prediction of a cluster that is truly prolate and inclined at an angle
$i\in[0, 90^\circ]$ (shown on the top left-most corner), under 4
deprojection scenarios: $D^p_{90}$ (second from left), $D^p_{min}$
(third from left), $D^o_{90}$ (second from right) and $D^o_{min}$
(right). The cluster is inclined such that its intrinsic axial ratio
$q$ results in the POS projected axial ratio of $b: a/q$. The
intrinsic semi-axes lengths that need to be ascribed to the 4
different $D$ configurations are: $\xi^p_{90} < \xi^T$, $\xi^p_{min} >
\xi^T$, $\xi^o_{90} < \xi^T$ and $\xi^o_{min} \approx \xi^T$. The
comparison in terms of the LOS extent is such that only for
$D^p_{min}$, it is in excess of that in the true system; for the other
3 $D$, it falls short of that in the true system. In light of these
comparisons, (according to the last bulletted list in Section~3.2.2),
the recovered deprojected density profile along $x$, is shown in the
lower panels (in solid lines), for the corresponding $D$, along with
the true density profile, plotted in broken lines. }}
\label{fig:prolate_fig}
\end{figure*}

A generic $T^p_i$ system is shown on the left in
Figure~\ref{fig:prolate_fig}; the X-ray $SB$ observed from this true
system is then deprojected under the 4 deprojection scenarios that are
shown in progression in Figure~\ref{fig:prolate_fig}. In the context
of this figure, the truly prolate system is projected along an
inclination such that the projected axes on the POS are $b$ ($< a$)
and $a/q$ (by construction of the coordinate systems). If we now
assume that this projection on the POS is due to a system that is
prolate and inclined at 90$^\circ$ to the LOS, then that system will
be ascribed intrinsic axes of $a/q, a/q \& b$. If however, the system
is assumed to be prolate and inclined at an angle $i_{min}$ ($<i$),
the long axis of the system will have to exceed that of the true
system, in order for this configuration to project to the
observations.  The state of the system, if assumed oblate and inclined
at either 90$^\circ$ or $i_{min}$, is shown in the second from right
and right-most positions in Figure~\ref{fig:prolate_fig}. 

Thus, we see that given a 2-D image, it is possible to guess the
semi-axes lengths of the 4 systems that can be described by the 4
different $D$ that we use herein. Now the question arises as to how
the run of $\rho$ with $x$ is dictated by the lengths of the semi-axes
in the corresponding configuration; more to the point, we try to gauge
how $\rho$ recovered for a given $D$ will compare to the that for
another choice of $D$, as well as the true system. The key to such
trends lies in the mutual weighing of two configurations for their LOS
extents and the ellipsoidal radii to a generic point.

For instance, given that the long axis of $D^p_{90}$ is shorter than
that of $T^p_{i}$, we realise that the ellipsoidal radius to a point
is higher in $T^p_{i}$ than in $D^p_{90}$ (2$^{nd}$ itemised point in
the last list in Section~\ref{sec:main}). Thus, the density profile
recovered by deprojecting under $D^p_{90}$ will have a similar shape
as the true $\rho$, but will be lower in amplitude. However, the
details of the 3 semi-axes lengths of the $D^p_{90}$ suggests that the
LOS extent is lower in $D^p_{90}$ than in $T^p_{i}$. Thus, the
lessening of the amplitude caused by differences in $\xi$ is
compensated by differences in LOS extent. Similarly, we can track how
the $\rho$ recovered for the other 3 $D$ will tally with the true
density profile, as well as how these recovered profiles compare with
each other. The details of the other cases are listed below. We then
register such an intercomparison as {\it possible}, when the true
system is $T^p_{i}$, with $i\in(i_{min}, 90^\circ$). Thus, whenever
the $SB$ of an observed system, when deprojected under the 4 used
$D$s, correspond to $\rho$ profiles that all have the same shape and
the amplitudes of which tally in the way described below, we turn our
experience around and declare the system to be prolate in nature with
inclination $<i_{min}$.

Had the inclination of the true system at hand ($T^p_i$) been such
that $i < i_{min}$, then even for $D^p_{min}$, the long-axis would
have fallen short of the long axis of the true system, as for the
$D^p_{90}$ case. In this case, the recovered profile of $\rho$ would
have compared with the true density profile similar to the $\rho$
recovered for $D^p_{90}$. Thus, the intercomparison of the density
profiles recovered in this case is registered as {\it possible} when 
the true system is $T^p_i$, with $i\in(0, i_{min}$).

We qualify the word ``possible'' in the two instances above, simply
because we are finding a possible answer to the question of: what is
the correct shape and inclination of the cluster that projects to the
observations. In other words, we do not claim that our answer is
unique. It maybe possible in some circumstances, to conjure up a
different geometry+inclination specification, that will correspond to
the same projection. We discuss this in Section~\ref{sec:discussions}.

Had we used other deprojection configurations, the intrinsic axial
ratios in such configurations would have been different from what they
are for the 4 $D$ that we use. The intercomparison of the profiles for
$\rho$ that emanate from such conjured configurations (the $D$ of our
choice), is rendered easier to interpret given that the $D$ of our
choice correspond to the extrema of the geometry and inclination
scales.

Some details:
\begin{itemize}
\item intrinsic axial ratio of this true system is higher than the
deprojection models inclined at 90$^\circ$, 
\item with the intrinsic major-axis exceeding that of the $D^p_{90}$
case but the intrinsic minor-axis exceeding that of the $D^o_{min}$
model;
\item the intrinsic major axis of the true system falls short of the same
in the $D^p_{min}$ but the intrinsic minor axis falls short of that in the
$D^o_{90}$ model.
\end{itemize}

For the 4 different deprojection models, the characteristics are
enumerated as follows:
\begin{enumerate}
\item $D^p_{90}$: 
\begin{itemize}
\item $LOS_D < LOS_T\Longrightarrow$ amplitude of $\rho_D$ at the
centre exceeds that of $\rho_T$.
\item Ellipsoidal radius to a general point $x,y,z$ is $\xi_D <
\xi_T\Longrightarrow\rho_D > \rho_T$.
\item This results in deprojected density profile ($\rho^x_D$) showing
a similar shape as $\rho^x_T$, but is lower in amplitude.
\end{itemize}
%
\item $D^p_{min}$:
\begin{itemize}
\item $LOS_D > LOS_T \Longrightarrow$ amplitude of $\rho_D$ falls
short of $\rho_T$.
\item $\xi_D > \xi_T$. 
\item The density comparison elicited by this inequality implies that
on the outside, $\rho^x_D$ is lower in amplitude and is also flatter
than $\rho^x_T$. This flatness will be less than that for a
similarly eccentric, true oblate system, for this deprojection
scenario (see next sub-section).
\end{itemize}
\item $D^o_{90}$:
\begin{itemize}
\item $LOS_D=a$, which is in excess of $LOS_T\Longrightarrow$
amplitude of $\rho_D$ falls short of that of $\rho_T$.
\item $\xi_D < \xi_T$. 
\item This implies that $\rho^x_D$ shows up with a shape that is
similar to the true system but is lower in amplitude; in fact the
recovered profile is similar to that recovered from $D^p_{min}$.
\end{itemize}
\item $D^o_{min}$:
\begin{itemize}
\item $LOS_D < LOS_T\Longrightarrow$ amplitude of $\rho_D > \rho_T$.
\item Given that this case corresponds to a smaller minor axis than
the true system, and that the true system is described by a
comparatively larger intrinsic major axis, $\xi_D \approx \xi_T$ for
locations that overlap between the true system and deprojection model.
\item $\rho_x(D^o_{90})$ has similar shape as the true system and is
highest in amplitude out of the 4 deprojection models.
\end{itemize}   
\end{enumerate}

\subsubsection{True system $T^o_i$}
\noindent
Following similar logic as used in the last section, we conclude that
when the observed cluster is oblate, inclined at an angle $i| i_{min}
< i < 90^\circ$, the density profiles recovered from the 4 deprojection
models are as follows.
\begin{itemize}
\item $D^p_{90}$: recovered density profile will have similar shape as
the true system but will be higher in amplitude from LOS considerations.
\item $D^p_{min}$: recovered density profile will manifest a larger
core than the true profile and higher in amplitude from LOS
considerations.
\item $D^o_{90}$: similar shape to true profile but amplitude will be
lower from considerations of ellipsoidal radius but can be either
lower or higher than the true profile, from LOS considerations.
\item $D^o_{min}$: recovered profile will have the same shape and
higher amplitude.
\end{itemize}

\begin{figure*}
\centering
\includegraphics[width=13cm]{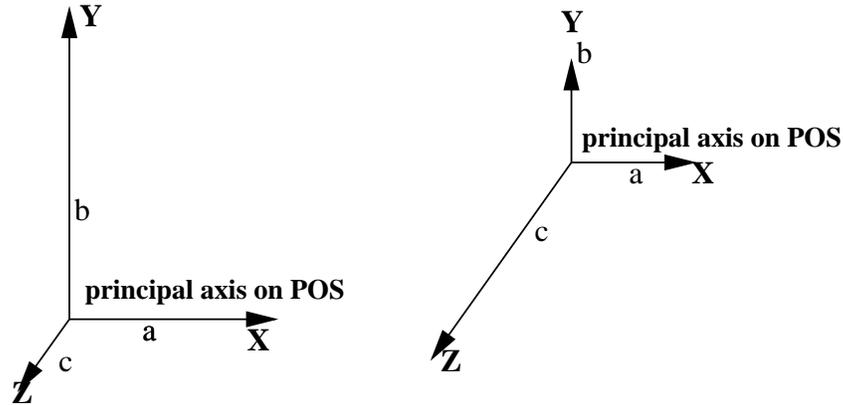}
\vskip-1.5cm
\caption{\small{The three kinds of triaxial systems encountered in
this work - Type~I and Type~II (left, $b>a>c$) and Type~III (right,
$b<a<c$). The high inclination and low $b/a$ ratio in Type~II systems
ensures that the photometric major axis is along the $x$-direction,
unlike in Type~I systems.}}
\label{fig:triaxial}
\end{figure*}

\subsubsection{Triaxial Systems}
\noindent
Here are some types of $triaxial$ systems that we can identify via the
inter-comparison of the deprojected luminosity density profiles in our
sample clusters. This is discussed in terms of the extent along the
principal axis that is identified as the LOS for an inclination of
90$^\circ$ - referred to as the $Y$-principal axis in the following
discussion.
\begin{itemize}
\item A triaxial configuration that deviates from a perfectly oblate
shape in that the $Y$-principal axis is the longest of the three while
the principal axis that corresponds to a photometric axis being the
second longest - this is referred to as ``triaxial-Type~I'' (left-most
panel in Figure~\ref{fig:triaxial}).

\item It is also possible that the true system is triaxial in the same
way as Type~I systems but the extent along the $X$-axis is only
slightly smaller than that along $Y$, so that for inclinations close
to 90$^\circ$, the photometric semi-axes in the $x$-direction exceeds
that along the $y$-direction, in contrast to the POS picture for
Type~I systems. Such a triaxial configuration is referred to as
``triaxial-Type~II'' (middle panel in Figure~\ref{fig:triaxial}).

\item If the sole deviation from pure prolateness that is responsible
for triaxiality is such that the $Y$-principal axis is the shortest of
the three, while the principal axis on the POS has intermediate
extent, then the ensuing triaxial configuration is referred to as
``triaxial-type~III'' (right panel of Figure~\ref{fig:triaxial}).

\item Since the relation between the projected and intrinsic axial
ratios in the triaxial case is different from that of the oblate and
prolate cases, it is not possible to extract accurate constraints on
the system inclination from the deprojection work alone. However, in
lieu of any other information, we can make an approximate guess about
what the inclination range for a certain triaxial cluster is. This is
done by suggesting if the inclination is intermediate to 90$^\circ$
and $i_{min}$ or less than $i_{min}$. The correct estimate of the
allowed inclination range for a triaxial cluster will be made on the
basis of the pair of relations (that are discussed below), between the
projected and intrinsic axial ratios of the triaxial system.

\end{itemize}

\subsubsection{Other Important Predictions}
\noindent
We can pursue logic similar to what was used above, to arrive at the
following, critically important conclusions.
\begin{itemize}

\item When the true system is $T^p_{90}$, all profiles will be similar
in shape, with the relative amplitudes of the profiles recovered under
the 4 different deprojection scenarios, as for a truly prolate system,
inclined at a general angle $i$.

\item When the true system is $T^o_{90}$, all profiles will be
similar, with the amplitudes as in the case of inclination at a general
angle $i$.

\item For a truly prolate cluster, the deprojected profiles will be
further apart, the smaller the angle of inclination is. On the other
hand, for a truly oblate cluster, the amplitude differences between
the recovered density profiles will increase with decreasing
inclination. 

\end{itemize}
%


\section{Tests}
\noindent
Here, we report the tests that were carried out to confirm the
predictions that were discussed in the last section. These tests were
performed with 3 clusters that are either perfectly prolate or oblate,
inclined at an angle intermediate to 90$^\circ$ and $i_{min}$, namely
60$^\circ$. In order to demonstrate the independence of the results
from the core radius of the system, 2 of the 3 test models were
assigned a core radius of approximately 0.01 arcmin and the other has
a core radius of about 0.2 arcmin; all the test clusters are assigned
a projected axial ratio of $q_p$=1.2. Surface brightness data for each
of these test models were estimated by performing a LOS integration of
the X-ray luminosity density distributions that were chosen to admit
an analytically integrable form.

Each of the test brightness profiles was deprojected under the 4
extreme deprojection scenarios: $D^p_{90}$, $D^p_{min}$, $D^o_{90}$
and $D^o_{min}$. The density profile that was recovered by
deprojecting the brightness data of any test model, under any
deprojection scenario, was re-projected back on the POS and this POS
distribution was then compared to the input brightness data, to ensure
the acceptability of the test run. 

\begin{figure*}
\vskip-5cm
\hskip-2cm
\includegraphics[width=18cm]{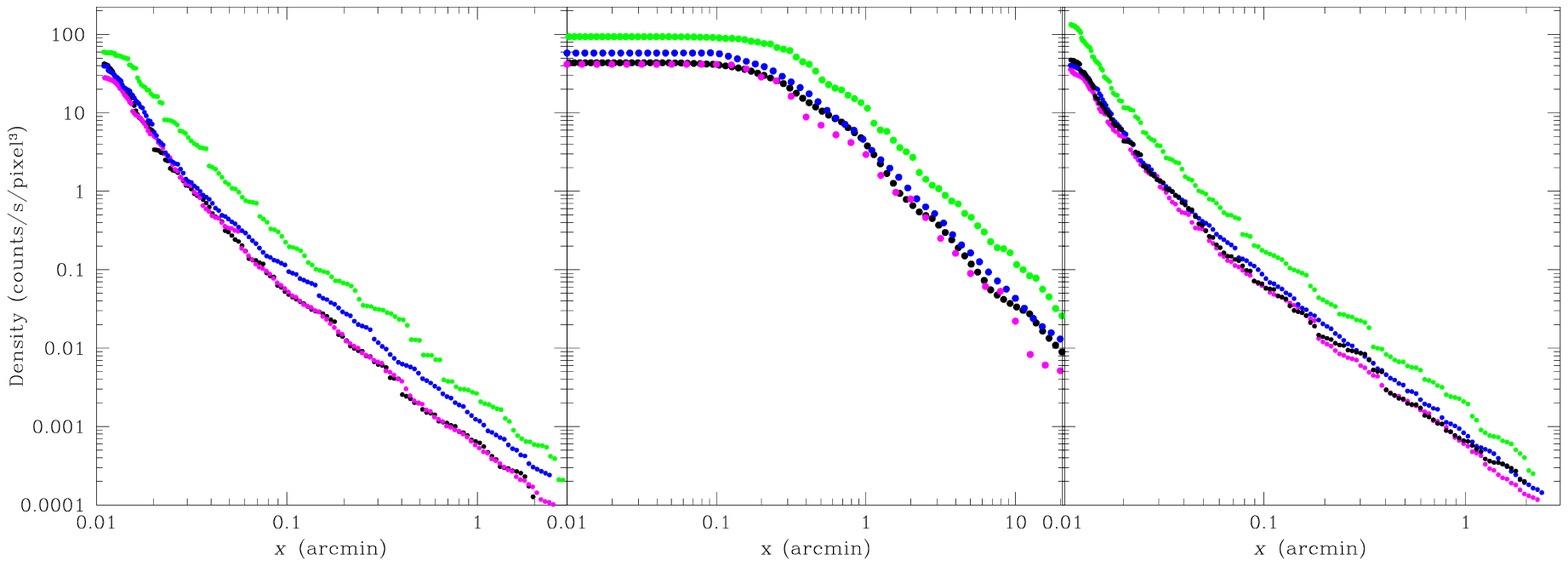}
\vspace{-14cm}
\caption{\small{X-ray luminosity density profiles of different test
clusters, recovered by deprojecting the toy X-ray surface brightness
maps of these systems. The test systems include prolate (right and
middle) and oblate (left) models with core size of about 0.003 arcmin
(left and right) and 0.2 arcmin (middle); the projected axial ratio is
1.2 for all three test systems. The density profiles recovered under
the $D^p_{90}$ scenario is in magenta, the $D^p_{min}$ case in blue,
$D^o_{90}$ in black and $D^o_{min}$ in green. }}
\label{fig:test}
\end{figure*}

The X-ray luminosity density profiles that were recovered with the
different deprojection scenarios, for the 3 different test models, are
displayed in Figure~\ref{fig:test}. The similarity of the
inter-comparison of the 4 different deprojected density profiles, in
the cases of the two prolate clusters with the varying core radii,
vindicates the core size independence of the shape determination.
However, is this an acceptable conclusion, in terms of the expanse and
details of the test systems we considered?

\subsection{Generality of Tests}
\noindent
In this section we examine the fundamental question of how general our
test systems are. To begin with, a statistical confirmation of our
method is a far cry from the limited tests that we have presented
below. It would be judicious to scan a range of ellipticities and core
radii corresponding to the distribution of these quantities, as
constrained by observations and simulations. Furthermore, the redshift
dependence of these distributions has to be folded in, to test for our
shape determination technique (as indicated by the ellipticity-redshift
relation). In fact, work is underway to calibrate our
technique against a sample of simulated gas+dark matter halos from the
Millennium Gas Simulations (Chakrabarty et. al, 2008).

While such a rigorous statistical test is on, we have presented here
tests done with three clusters that have characteristics similar to
what have been noted with the sample of clusters that we apply our
method to (following section) - as apparent from Table~1, in our
cluster sample, there are 7 out of 25 clusters, that have a projected
axial ratio lying between 1.2 and 1.25, approximately (including A~399
which is at a redshift of 0.072). Most of the other systems manifest
even more eccentric 2-D images. In fact, \cite{betty05} state that
the redshift-ellipticity relation is not manifest in this
sample. Thus, our test configurations as motivated by measurement.

\section{Application}
\noindent
We apply our formalism to a sample of 25 clusters
from~\cite{reese02}. This sample consists of 18 X-ray selected clusters
with $z\geq 0.14$, $\delta\geq -15\degr$ and $L_{\rm X}(0.1 - 2.4\
{\rm keV})\geq 5 {\times} 10^{44}\ h_{50}^{-2}\ {\rm erg\ s}^{-1}$ and
for which high S/N detections of SZE, high-S/N X-ray imaging and
electron temperatures were available.  To these, $7$ clusters from
the~\cite{mason01} sample were added, containing clusters from X-ray
flux-limited catalogue of~\cite{ebe96}. Details on the completeness of
the latter subsample are given by~\cite{mason00}. Actually, we choose to
work with 24 of these 25 clusters since the measured surface
brightness profile of one of the sample clusters (A~520) appears to be
plagued by very large observational error bars in the last 3 arc
minutes (typically, in excess of 50$\%$). 

The motivation behind choosing this sample is that the
three-dimensional intrinsic morphology of the clusters in this sample
has already been reported, as estimated by an independent method that
combines X-ray and Sunyaev-Zel'dovich
observations~\citep{sereno06}. Thus, we will be in a position to
compare our results with that from \cite{sereno06}.

\subsection{Data Analysis}
In our analysis we have used archival Chandra or XMM-Newton data for
all clusters in the sample. The Chandra Interactive Analysis of
Observation software (CIAO 3.3.0.1) and the XMM-Newton Science
Analysis Software (SAS 7.0.0) were used throughout.

We first modelled the projected two dimensional cluster emission, in
the $0.3-7.0\ {\rm keV}$ energy band. Pixel values of all detected
point sources were replaced with values interpolated from the
surrounding background regions. Using the {\it SHERPA} software, we
fitted the cluster surface brightness to elliptical 2D
$\beta$-models. The resulting best fit parameters for all clusters in
the sample, relevant for this paper, are listed in
Table~\ref{tab:2Dfit}.

We then extracted one dimensional surface brightness profiles within
elliptical concentric annuli, with projected ellipticity, position
angle and centre matching those estimated in the our 2D analysis (as
listed in Table~\ref{tab:2Dfit}). The distance $x$ from the cluster
centre is measured along the major axis of the ellipses. The profiles
were extracted in the $0.3-7.0\ {\rm keV}$ energy range. In the same
energy band, the background for each cluster was estimated, within a
peripheral region of the field of view. Corresponding radial profiles,
extracted within the same elliptical regions from the exposure maps,
provided the required average values of the effective exposure time
within each annulus. Background subtracted and exposure corrected
surface brightness profiles were then extracted for each cluster.

\begin{table*}
\caption{Cluster Sample}
\centering
\begin{tabular}{lcccccc}
\hline\hline
 &  & \multicolumn{2}{c}{$x_{\rm c},y_{\rm c}$}  &   &  & \\
 &z & R.A.&Decl. & $q$ & $\theta$& Satellite\\
 &    &      &  &	         & deg 	   & 	\\
\hline\hline
MS 1137.5+6625  &$0.784$  &$11\ 40\ 22.3$ & $+66\ 08\ 15.3$  &$1.113\pm0.014$ & $116.1\pm1.0$&   $1$\\
MS 0451.6-0305  &$0.550$  &$04\ 54\ 11.4$ & $-03\ 00\ 51.3$  &$1.307\pm0.015$ & $95.9\pm1.1$ &   $1$\\
Cl 0016+1609    &$0.546$  &$00\ 18\ 33.5$ & $+16\ 26\ 12.9$  &$1.205\pm0.013$ & $310.8\pm1.7$&   $2$\\
RXJ1347.5-1145  &$0.451$  &$13\ 47\ 30.7$ & $-11\ 45\ 09.1$  &$1.453\pm0.019$ & $158.5\pm1.0$&   $1$\\
A 370           &$0.374$  &$02\ 39\ 53.3$ & $-01\ 34\ 39.0$  &$1.564\pm0.018$ & $6.2\pm0.7$  &   $1$\\
MS 1358.4+6245  &$0.327$  &$13\ 59\ 50.7$ & $+62\ 31\ 04.1$  &$1.325\pm0.019$ & $156.6\pm1.4$&   $1$\\
A 1995          &$0.322$  &$14\ 52\ 57.9$ & $+58\ 02\ 55.8$  &$1.242\pm0.010$ & $57.8\pm1.0$ &   $1$\\
A 611           &$0.288$  &$08\ 00\ 56.8$ & $+36\ 03\ 23.5$  &$1.14\pm0.05$   & $34\pm9$     &   $1$\\
A 697           &$0.282$  &$08\ 42\ 57.6$ & $+36\ 21\ 56.8$  &$1.334\pm0.016$ & $163.8\pm1.2$&   $1$\\
A 1835          &$0.252$  &$14\ 01\ 02.0$ & $+02\ 52\ 42.9$  &$1.225\pm0.012$ & $173.0\pm1.4$&   $1$\\
A 2261          &$0.224$  &$17\ 22\ 27.1$ & $+32\ 07\ 57.4$  &$1.022\pm0.017$ & $90.0\pm1.7$ &   $1$\\
A 773           &$0.216$  &$09\ 17\ 53.1$ & $+51\ 43\ 37.9$  &$1.237\pm0.022$ & $90.0\pm2.5$ &   $1$\\
A 2163          &$0.202$  &$16\ 15\ 46.6$ & $-06\ 08\ 44.9$  &$1.206\pm0.004$ & $90.0\pm0.6$ &   $1$\\
A 1689          &$0.183$  &$13\ 11\ 29.6$ & $-01\ 20\ 28.0$  &$1.141\pm0.012$ & $17.6\pm2.2$ &   $1$\\
A 665           &$0.182$  &$08\ 30\ 57.1$ & $+65\ 51\ 01.8$  &$1.238\pm0.012$ & $146.3\pm1.2$&   $1$\\
A 2218          &$0.171$  &$16\ 35\ 51.9$ & $+66\ 12\ 34.6$  &$1.162\pm0.009$ & $96.5\pm1.5$ &   $1$\\
A 1413          &$0.142$  &$11\ 55\ 17.9$ & $+23\ 24\ 16.2$  &$1.473\pm0.019$ & $177.8\pm0.9$&   $1$\\
A 2142          &$0.091$  &$15\ 58\ 20.1$ & $+27\ 14\ 03.5$  &$1.540\pm0.007$ & $127.7\pm0.3$&   $1$\\
A 478           &$0.088$  &$04\ 13\ 25.3$ & $+10\ 27\ 53.5$  &$1.477\pm0.006$ & $43.5\pm0.3$ &   $1$\\
A 1651          &$0.084$  &$12\ 59\ 21.9$ & $-04\ 11\ 44.6$  &$1.184\pm0.013$ & $87.6\pm1.9$ &   $1$\\
A 401           &$0.074$  &$02\ 58\ 57.1$ & $+13\ 34\ 37.8$  &$1.303\pm0.008$ & $34.9\pm0.6$ &   $1$\\
A 399           &$0.072$  &$02\ 57\ 52.0$ & $+13\ 02\ 38.7$  &$1.207\pm0.009$ & $22.5\pm1.2$ &   $2$\\
A 2256          &$0.058$  &$17\ 04\ 00.4$ & $+78\ 38\ 37.1$  &$1.327\pm0.008$ & $28.7\pm0.5$ &   $2$\\
A 1656    	&$0.023$  &$12\ 59\ 44.1$ & $+27\ 56\ 43.0$  &$1.141\pm0.006$ & $90.0\pm0.6$ &   $2$\\
\hline
\label{tab:2Dfit}
\end{tabular}
\begin{list}{}{}
\item Col. 1: Cluster name. Col. 2: Cluster redshift. Cols.3-6: fit
parameters of the elliptical 2D $\beta$ model: $x_{\rm c},y_{\rm c}$
is the central position; $q$ is the projected axial ratio, and
$\theta$ is the orientation angle (north over east). In the last
column, label $1$ is for Chandra and $2$ for XMM observations.
\end{list}
\end{table*}

\section{Results}
\noindent
The intrinsic shapes and inclination classes that we infer for our
chosen sample of clusters are enumerated in Table~2. We also include
the shapes and inclinations of the clusters, as predicted by
\cite{sereno06}, in adjacent columns.

\begin{figure*}
\centering
\includegraphics[width=15cm]{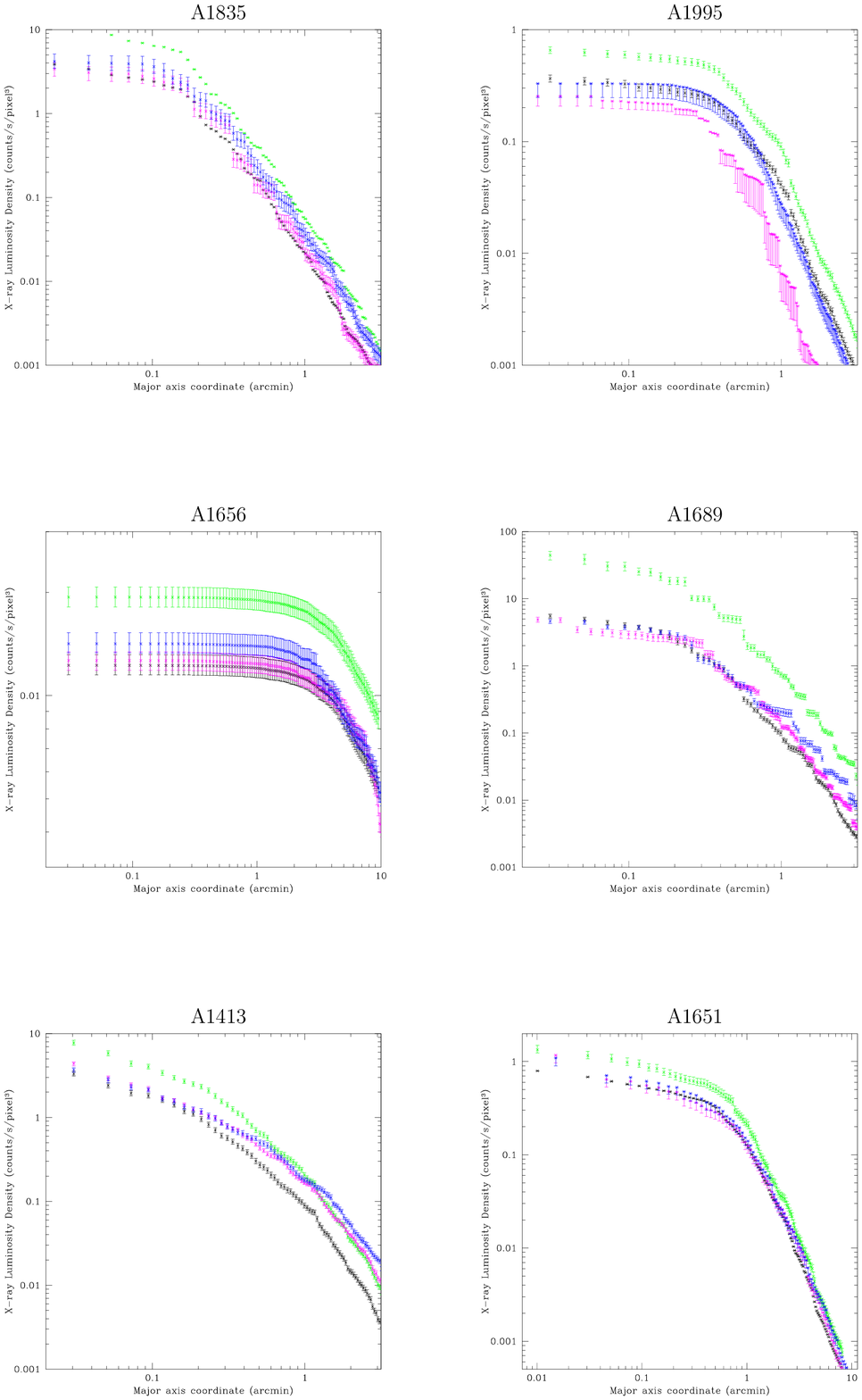}
\caption{\small{X-ray luminosity density profiles of 6 of the 24
clusters in our sample. The colour scheme used for the profiles
recovered under the 4 different deprojection scenarios is the same as
in Figure~\ref{fig:test}.  }}
\label{fig:profiles}
\end{figure*}

\begin{table*}
\caption{Shapes $\&$ Inclinations of Cluster Sample}
\centering
\begin{tabular}{lcccc}
\hline\hline
{} & Inferred Shape  & Approximate Incl  & Shape in \cite{sereno06} & Incl in \cite{sereno06} \\
{} &                 & (degrees)         &                         & (degrees) \\
\hline\hline
MS 1137.5+6625  & prolate &  $26<i<90$ & prolate & 16$\pm$9\\
MS 0451.6-0305  & triaxial-Type~I & $40<i<90$ &prolate/oblate & 56$\pm$13/58$\pm$43\\ 
Cl 0016+1609    & triaxial-Type~I & $34<i<90$ &prolate &26$\pm$10\\
RXJ1347.5-1145  & triaxial-Type~III & $i<47$ &oblate/prolate &84$\pm$117/35$\pm$12\\
A 370           & triaxial-Type~II & $i<50$ &prolate & 27$\pm$12\\
MS 1358.4+6245  & prolate & $i<41$ &oblate & 52$\pm$8 \\
A 1995          & triaxial-Type~III & $i<36$ &oblate/prolate & 64$\pm$35/46$\pm$27\\ 
A 611           & prolate & $29<i<90$ &prolate &35$\pm$32\\
A 697           & triaxial-Type~I & $42<i<90$ & prolate& 26$\pm$10\\ 
A 1835          & prolate & $35<i<90$ & prolate& 29$\pm$7 \\
A 2261          & prolate & $i<12$ & prolate& 10$\pm$6 \\
A 773           & triaxial-Type~II & $i<36$ & prolate &14$\pm$5\\  
A 2163          & triaxial-Type~I & $34<i<90$ &prolate & 28$\pm$10\\ 
A 1689          & prolate & $29<i<90$ &oblate/prolate & 65$\pm$44/46$\pm$23\\ 
A 665           & triaxial-Type~I & $36<i<90$ &oblate &47$\pm$13  \\
A 2218          & triaxial-Type~I & $34<i<90$ &prolate &22$\pm$11  \\
A 1413          & triaxial-Type~II & $i<47$ &oblate/prolate&73$\pm$33/37$\pm$15\\ 
A 2142          & triaxial-Type~II & $i<50$ &oblate/prolate&79$\pm$42/34$\pm$10 \\
A 478           & triaxial-Type~II & $i<48$ & prolate & 23$\pm$12 \\   
A 1651          & prolate & $i<32$ &prolate & 12$\pm$7\\
A 401           & triaxial-Type~II & $i<40$ & prolate& 25$\pm$6\\ 
A 399           & oblate & $34<i<90$ & oblate& 38$\pm$5 \\
A 2256          & triaxial-Type~II & $i<41$ &prolate &34$\pm$15\\  
A 1656    	& prolate & $i<29$ &prolate &33$\pm$30  \\
\hline
\end{tabular}
\begin{list}{}{}
\item Col. 1: Cluster name, Col. 2: Intrinsic cluster shape recovered
from our work, Col. 3: Recovered inclination class - in degrees,
Col. 4: Shape reported by Sereno et. al (2006) $\&$ Col. 5:
Inclination as reported by Sereno et. al (2006) (in degrees) - the
clusters for which both prolate and oblate solutions were found
admissible by \cite{sereno06} are marked accordingly in Col~4 and the
corresponding inferred inclinations are shown in Col~5.
\end{list}
\end{table*}

Thus, we find that in our sample of 24 clusters, 15 are triaxial, 1
is oblate and the rest (8) are prolate. This shape distribution is
significantly different from that deciphered by \cite{sereno06}, as
evident in Table~2; we discuss the differences in detail in
Section~\ref{sec:diff}.

We also looked at the central flux distribution for the clusters in
our sample and sought correlations with the recovered shapes; in
particular, we attempted to establish is any correlation exists
between the clusters and the central flux measurements. This was found
to be absent.

\subsection{Incorporation of SZe Data to Prolate $\&$ Oblate Clusters}
\noindent
The impending improvement in the quality of the SZe data will boost
this technique greatly. Knowledge of the extent along the LOS will
permit us to perform the deprojection in a fully triaxial geometry -
such a deprojection is possible with DOPING. Thus, while in the
current version of DOPING we compare the plane-of-the-sky projection
of the recovered density distribution to the observed brightness map
of the system, on improvement of the SZe data, the LOS extent of the
density structure can also be pinned down, for a given choice of the
inclination. We would then be in a position to better constrain both
the intrinsic axial ratios for the triaxial case and thereby improve
upon the estimate of the inclination of the cluster to the LOS.

As mentioned above, one motivation for choosing this sample is the
availability of the SZe data for all the clusters in the sample. This
was reported by \cite{betty05}. It is not surprising that the ratio of
the LOS extent to the photometric major axis ($q_{LOS}$) of a sample
cluster is characterised by large error bars. 

We implement the reported $q_{LOS}$ first for all the clusters that
are recovered as prolate in our sample, and then for the oblate
ones. To do this, we use the following relations (from
\cite{sereno06}):
\begin{eqnarray}
q_{LOS} &= \displaystyle{\frac{\sqrt{q_p^2 - \cos^2{i}}}{q_p^2\sin{i}}} \quad{\rm prolate} \\ \nonumber
{} &= \displaystyle{\frac{\sqrt{1 - q_p^2\cos^2{i}}}{\sin{i}}} \quad{\rm oblate}
\end{eqnarray}

The reported $q_p$ and $q_{LOS}$ values are used in this equation, for
a sample cluster that has been identified as prolate. The values of
{\it $q_{LOS}$ reported at the two extrema of the error bands, for any
cluster, results in a range for the inclination}.  These inferred
range on the inclination is superimposed on the range of inclinations
that were judged admissible from deprojection considerations
alone. The range corresponding to the overlap of these two sets of
constraints is then narrower, as represented in the last column of
Table~3. A similar exercise was undertaken for the sole oblate cluster
in our sample (according to our shape evaluation); the improved
inclination estimate for this system is included as the last entry in
Table~3.

As can be appreciated from this table, the inclusion of SZe data makes
a significant effect in constraining the cluster inclination. However,
the allowed inclination ranges from the X-ray data and the SZe data
did not overlap for the case of Abell 611; thus, we are unable to produce 
an inclination for this cluster.

\begin{table*}
\caption{Improved Inclinations (using SZe Data) of the Prolate and
Oblate Sample Clusters} \centering
\begin{tabular}{lcccc}
\hline\hline
{} & Inferred Shape  & Incl from Deprojection & from SZe data & Refined Incl\\
{} &                 & (degrees)              & (degrees) &(degrees)\\
\hline\hline
MS 1137.5+6625  & prolate &  $26<i<90$ & $11<i<63$ & $26<i<63$\\
MS 1358.4+6245  & prolate & $i<41$ & $0<i<44$ & $41<i<44$ \\
A 611           & prolate & $29<i<90$ & $0<i<20.665$ & $-$\\
A 2261          & prolate & $i<12$ & $7<i<20$ & $7<i<12$ \\
A 1689          & prolate & $29<i<90$ & $0<i<33$ & $29< i<33$\\
A 1651          & prolate & $i<32$ & $8<i<32$ & $8<i<32$\\ 
A 1656    	& prolate & $i<29$ & $19<i<34$ & $19<i<29$\\ 
A 1835          & prolate & $35<i<90$ & $13<i<78$ & $35<i<78$ \\
A 399           & oblate & $34<i<90$ & $34<i<37$ & $34<i<37$ \\
\hline
\end{tabular}
\begin{list}{}{}
\item Col. 1: Cluster name, Col. 2: Intrinsic cluster shape recovered
from our work, Col. 3: Inclination recovered from the deprojection
exercise, Col. 4: Inclination as indicated by relating the projected
POS axial ratio and LOS extent \citep{betty05}, Col. 5:
The most improved bounds on inclination, obtained as overlap of
constraints imposed by the deprojection exercise and the SZe data.
\end{list}
\end{table*}

\subsection{Incorporation of SZe Data to Triaxial Clusters}
\noindent
The direct exploitation of the SZe data is possible only for the cases
of the purely oblate or purely prolate clusters. For the general
triaxial clusters, the SZe information is used to process the two
intrinsic axial ratios, given that we know how the lengths of the
three semi-axes compare, given the recovered triaxiality
type. Additionally, the cluster inclination can be pinned down from
the recovered inequalities that relate the intrinsic axial ratios to
unity. Such constraints too are derived from the identification of the
type of triaxiality in question. The details of the methodology used
to constrain axial ratios and inclinations is delineated in Appendix
A.

As mentioned above, the inclinations guessed from the X-ray
deprojection exercise alone are loose approximations only and only the
values of $i$ derived from the incorporation of the SZe data should be
considered correct.

The axial ratios in the considered triaxial cases are tabulated, along
with the improved inclination constraints, in Table~4. It is to be
noted that these constraints are estimated by ignoring the errors that
are reported \citep{betty05} on the elongation of the sample
clusters. The axial ratios are calculated with the inclination set to
the median of the band of inclinations that are deemed suitable for a
given cluster.

\begin{table*}
\caption{Axial Ratios and Improved Inclinations (using SZe Data) of
the Triaxial Sample Clusters} \centering
\begin{tabular}{lccccc}
\hline\hline
{} & Shape from Deprojection & Approx Incl from Deproj & Correct Incl & $q_1$ & $q_2$\\
{} &                         & (degrees)               &(degrees) &&\\
\hline\hline
MS 0451.6-0305  & triaxial-Type~I & $40<i<90$ & $67<i<90$ & 0.92 & 1.34\\
Cl 0016+1609    & triaxial-Type~I & $34<i<90$ & $73<i<90$ & 0.65 & 1.24\\
A 697           & triaxial-Type~I & $42<i<90$ & $77<i<90$ & 0.58 & 1.33\\
A 2163          & triaxial-Type~I & $34<i<90$ & $72<i<90$ & 0.68 & 1.24\\
A 665$^{*}$     & triaxial-Type~I & $36<i<90$ & $70<i<90$ & 0.78 & 1.27\\
A 2218          & triaxial-Type~I & $34<i<90$ & $78<i<90$ & 0.45 & 1.20\\
A 370           & triaxial-Type~II & $50<i<90$ & $73<i<90$ & 0.84 & 1.62\\
A 773           & triaxial-Type~II & $36<i<90$ & $78<i<90$ & 0.48 & 1.28\\
A 1413          & triaxial-Type~II & $47<i<90$ & $66<i<71$ & 0.96 & 1.64\\
A 2142          & triaxial-Type~II & $50<i<90$ & $69<i<77$ & 0.96 & 1.67\\
A 478           & triaxial-Type~II & $48<i<90$ & $75<i<90$ & 0.73 & 1.52\\
A 401           & triaxial-Type~II & $40<i<90$ & $72<i<90$ & 0.76 & 1.34\\ 
A 2256          & triaxial-Type~II & $41<i<90$ & $67<i<90$ & 0.94 & 1.36\\
A 1995          & triaxial-Type~III & $i<36$ & $0<i<31$ & 3.94 & 0.34\\
RXJ1347.5-1145  & triaxial-Type~III & $i<47$ & $0<i<65$ & 5.14 & 0.29\\
\hline
\end{tabular}
\begin{list}{}{}
\item Col. 1: Cluster name, Col. 2: Intrinsic cluster shape recovered
from our work, Col. 3: Approximate inclination range, guessed from the
deprojection exercise, Col. 4: Bounds on inclination, obtained from
the constraints on the intrinsic axial ratios, as indicated by the
type of triaxiality (Type~I, II or III) that is assigned to the
cluster from deprojection work, Col. 5: Ratio $q_1$ between the length
of the principal axis ($a$) that corresponds to one of the photometric
axes and that along the $Y$-principal axis, ($b$, see
Figure~\ref{fig:triaxial}) Col. 6: Ratio ($q_2$) between $a$ and and
the other remaining principal axis ($c$).
\item $^{*}$ For this cluster, the $q_{LOS}$ value reported at the
higher end of the range is used; the medial value of $q_{LOS}$
constrains the cluster inclination to less than a degree.
\end{list}
\end{table*}

\section{Discussions}
\label{sec:discussions}
\noindent
In this work we have presented a simple but novel technique to extract
the intrinsic shape and inclination class of galaxy clusters, using
X-ray brightness maps alone. The availability of SZe data is then
shown to improve the inclination estimate considerably.  The main
motivation behind this exercise is the improvement on mass estimates
which is expected to refine the constraints that we can place on
cosmological constants, from the analysis of cluster data.

Our suggested formalism relies upon the inter-comparison of the
density profiles that are deprojected from the measured X-ray surface
brightness maps, under 4 different deprojection scenarios that combine
the extrema of the inclination scale and the geometry scale, when
clusters are treated as figures of revolution. It is found that each
type of observed system leaves its unique signature in this
inter-comparison amongst the deprojected X-ray luminosity density
profiles. The knowledge of the cluster elongation along the LOS (from
SZe data) is then used to supplement our shape and inclination
determination; in particular, the recovered constraints on the
inclination are significantly improved, if we believe the SZe
measurements. In case of triaxial systems, the intrinsic axial ratios
can also be tracked, using the SZe data.

\subsection{Differences with Results of \cite{sereno06}}
\label{sec:diff}
\noindent
Our shape determination agrees with the shape advanced by
\cite{sereno06}, for 13 of the 24 sample clusters. This
quantification includes those cases that we identify as triaxial and
\cite{sereno06} find compatible with both the prolate and oblate
geometries. For another 10 cases, while we find the system to be
definitely triaxial, \citep{sereno06} report the system to be
prolate. This might result from the fact that \cite{sereno06} use the
observed values of projected axial ratio and LOS extent, in equations
relating these quantities and the intrinsic axial ratio to inclination
$i$, for the two geometries of prolateness and oblateness (see
Equations~8-11 in Sereno et. al, 2006). 

They constrain their choice of geometry by identifying realistic
solutions for inclinations, such as $-1\leq\cos(i)\leq 1$. Thus, if
for example, they identify an unrealistic solution for $i$ for the
oblate case, they declare the system prolate. But strictly speaking,
all that they can conclude is that the cluster at hand is {\it not
oblate}. In other words, it is possible that the cluster is triaxial,
in such a way that given the high measurement errors, the equation for
$i$ in the prolate case also yields acceptable values. The 10 sample
clusters that we spot as triaxial, but \cite{sereno06} call prolate,
are similarly 10 non-oblate clusters. 

The one system that we identify as prolate and \cite{sereno06} call
oblate i.e. non-prolate, is MS~1358.4+6245. However, when we use the
values of 1.325 for the projected axial ratio (ignoring the $\pm$1$\%$
error) and 0.91 for $e_{LOS}$ (using the upper bound from the measured
range of 0.72$\pm$0.19), in Equations~8 and 9, we do actually get a
realistic inclination of about 44$^\circ$, under the assumption of
prolateness. In other words, the methodology used by \cite{sereno06},
does not rule out prolateness for this cluster, in line with our
inference.

Thus, our methodology is relatively more powerful, since we have
greater resolution ability than simply ruling out
prolateness/oblateness. Moreover, our implementation of the SZe data
allows for the recovery of a much narrower range of inclinations, for
the same measurement as used by previous workers.

\subsection{X-rays vs. Optical $SB$}
\noindent
We would like to emphasise that it is very much possible to utilise
the optical surface brightness distribution of clusters, if available,
to make an estimate of the morphological structure of the cluster with
the aid of DOPING.  This would allow the verification and
quantification of possible discrepancies between results obtained
using X-ray and optical data. This is particularly interesting, given
the study presented by \cite{gottboler_07} which suggests a relatively
more spherical central gas component, compared to the dark matter one.

To this aim, we are now applying the DOPING algorithm to both
dark matter halos from the Millennium Simulations and to gas+dark matter
 $SB$
distribution of the same halos (from the Millennium Gas Simulations).
Results will be published in Chakrabarty et. al, (2008). This
will answer the question of how the shape determination is affected by
 changes in the nature of the input.

In this connection, it is further important to mention that our sample
of real clusters show more eccentric gas distribution in projection,
than the simulated ones studied in the hydrodynamics simulations of
\cite{gottboler_07} and, consequently, the intrinsic axial ratios that
we recover, are not akin to $q_1$=1, $q_2=$2, as suggested in that
work.

\subsection{Core Sizes}
\noindent 
We tested our predictions on a suite of model clusters and found that
the recovery of shapes and inclinations is independent of the axial
ratios that were assigned to these test systems, as well as the core
sizes that these test systems were described by. That the core size is
not influential in the recovery of the intrinsic shape of the cluster
is not surprising since it is the outer part of the brightness profile
that is conventionally considered when identifying shapes of clusters.
Thus, whether it is Abell 2261 or MS 1358, our formalism is able to do
justice to systems at both ends of core sizes in the sample used
herein. This brings us to an important point in regard to the
applicability of our technique - it can be implemented to estimate the
3-D geometry of clusters at varying redshifts! In fact, the sample
that we used to illustrate the efficacy of our formalism is rather
eclectic in its redshift coverage, ($z$=0.023 to 0.784).

Crucially, our work has clearly indicated {\it the dependence of the
recovered core size of the deprojected luminosity density profile, on
the deprojection scenario} used. Thus, for Abell~665, the deprojected
density corresponding to an assumption of oblateness and $i=i_{min}$
implies a core size that is less than half ($\approx0.06$ arcmin)
that obtained by deprojection performed under prolateness and
$i=i_{min}$ and inclination combinations ($\approx$0.15 arcmin).

It is important to stress here that we {\it do not need to delete the
cores from our analysis} due to any conceived inability of our code to
deal with the modelling of the very central regions of the
clusters. The code used in the work is in fact non-parametric and its
functionality is not challenged by local changes in slopes of the
observed X-ray brightness profiles or even by local enhancements of
observational errors in the brightness distributions. The deprojected
profile will of course bear larger error bars when the input
brightness profile manifests the same, than otherwise. The fact that
the inclusion of the core does not affect the shape determination is
brought home by the tests that we have performed by varying the core
sizes of toy clusters.

\subsection{Central Density}
\noindent
One ramification of the observed nature of the luminosity-temperature
relationship in clusters is the need to invoke some (unknown) pressure
that is effective in reducing the density of the intracluster gas in
the cores \citep{voit02}. However, as we have seen in this work, the
X-ray luminosity density at the centres of clusters depends critically
on the deprojection scenario adopted in the model. Thus, in
Abell~1689, we notice that the density recovered under an assumption
of prolateness and $i=i_min$, at about 3 arcmin, is about 10 times
less than that recovered under the other three deprojection
scenarios. Thus, our work indicates the important contribution that
deprojection uncertainties can make towards the observed trends in the
cluster self-similar relationships.

\subsection{Effect of Errors of Cluster Elongation}
\noindent
The current state of affairs regarding the quality of SZe data is
indeed unsatisfactory, albeit improvements are impending. Thus, when
we use the SZe data in this work, we make the conscious decision to
work with the value of the cluster elongation that is reported at the
centre of the error band. Incorporation of the measurement errors
would have greatly reduced the quality of the constraints on the
inclinations.

\subsection{The Case of Abell~1995 and RXJ1347}
\noindent
The clusters Abell~1995 and RXJ1347 have been found to be triaxial, of
Type~III, which implies that these clusters are nearly prolate except
that $c < a$; also, we expect $b-a \ll a$. The density profiles
recovered under the assumption of prolateness are flatter for these
clusters than the profiles deprojected under oblateness. This
comparison is allowed if the ellipsoidal radius in the prolate cases,
in general, exceed that in the oblate cases. This is in turn ensured
if $a$ is the photometric semi-axis and if the inclination is
small. We find that the inclination is indeed constrained to very
small angles, given the reported values of $q_p$ and $q_{LOS}$ (rather
the medial value of these quantities within the reported measurement
error bands). Thus, these two clusters are deciphered to be nearly
face-on systems.

\subsection{The Case of Abell~370}
\noindent
The cluster Abell~370 was reported to bear a ``pronounced triaxial
morphology'' by \cite{betty05_370}, on the basis of the X-ray and SZe
data. However, \cite{sereno06} find the prolate solution to be consistent
with the observed X-ray and SZe data for this cluster. In contrast to
this, we actually find this cluster to be triaxial, with an
inclination in excess of 50$^\circ$.

\subsection{The Case of CL~0016+609}
\noindent
\cite{hughes98} suggested a ``reasonable triaxial'' morphology for the
cluster CL~0016+609, on the basis of the discrepancy between the
Hubble constant values that were estimated from the oblate and prolate
models. This suggestion is in line with our conclusion of a triaxial
shape for this cluster.

\subsection{Future Work}
The identification of the intrinsic shape and inclination of a cluster
is a major step in the characterisation of clusters, both in terms of
cluster masses and quantification of the contribution of deprojection
uncertainties to the observed scatter in the self-similar
relationships of clusters. These applications are planned in future
contributions. 

In fact, it is envisaged that the improved understanding of cluster
spatial configurations leads to better constraints on cluster
masses. If this is supplemented with information of mass distributions
from dynamical considerations (using an algorithm such as CHASSIS -
Chakrabarty $\&$ Saha, 2001), we could potentially place bounds on the
distribution of gas that is at hydrostatic equilibrium. Furthermore,
using estimation of the mass within a given projected radius from
lensing measurements, would constrain the dynamical mass distribution
even better, leading to added improvements in cluster
characterisation.  An exercise in the determination of improvement in
the dynamical mass of a lensing galaxy, using such lensing
constraints, is currently underway (Chakrabarty, et. al, 2008).

\begin{acknowledgements}
\noindent
DC is supported by a Royal Society Dorothy Hodgkin Fellowship. HR
would like to acknowledge the Mary Cannell Summer Studentship that
made her contribution possible.
\end{acknowledgements}

\vspace{1cm}
\noindent

\bibliographystyle{aa}

\appendix

\section[Improving on Inclination Constraints for Oblate $\&$ Prolate Systems]
\noindent
Given that in our coordinate system, $X=x$, we say:
\begin{equation}
(y, z) = {\bf R}_i (Y, Z)
\end{equation}
where ${\bf R}_i$ is the rotational matrix corresponding to a rotation
through the angle $i$. Let us consider a triaxial system, with
intrinsic axial ratios $q_1$ and $q_2$, i.e. ratios of the maximum
extent along the $X$-axis to that along the $Y$-axis is $q_1$ and the
ratio of the principle axis along the $X$-axis to that along the
$Z$-axis is $q_2$. 

In this triaxial configuration, the square of the ellipsoidal radius is
\begin{eqnarray}
\xi &=&\displaystyle{x^2 + {y^2}[q_1^2\cos^2(i) + q_2^2\sin^2(i)]} \\ \nonumber
    &{}&\displaystyle{+ {z^2}[q_1^2\sin^2(i) + q_2^2\cos^2(i)] 
			+ yz\sin(2i)(q_1^2 - q_2^2)}
\end{eqnarray}
This implies that the plane-of-the-sky projection $I(x,y)$ of any
intrinsic quantity $G(\xi)$, can be represented as:
\begin{eqnarray}
I(x,y) &=& \int_0^\infty{G(\xi)dz} \\ \nonumber
{} &=& \displaystyle{\int^\infty_\eta{G(\xi)\frac{d\xi}{\sqrt{\xi -\eta^2}}}}
\end{eqnarray}
where
\begin{equation}
\eta = \displaystyle{x^2 + y^2\frac{q_1^2{q_2^2}}{R^2}} 
\end{equation}
The notation here is chosen to concur with that in \cite{fabricant85}.
Since the RHS of this equation is a function of $\eta$ only and the
projected quantity $I(x,y)$ is a constant along the isophotes, the
equation relating $\eta$ to $x$ and $y$ must give the form of the
isophotes. Then the ratio $q_p$, of the extent along the $x$-axis to
that along the $y$-axis is related to ratio of the semi-axes along $X,
Y \& Z$-axes respectively as:
\begin{equation}
\displaystyle{q_p^2} = \displaystyle{\frac{{q_1^2}q_2^2}
                                          {q_2^2\cos^2(i) + q_1^2\sin^2(i)}
                                    }. 
\label{eqn:ep}
\end{equation}
A similar exercise involving integration over $y$ yields the ratio
$e_{LOS}$ between the extent along the $z$-axis and the $x$-axis as:
\begin{equation}
\displaystyle{e_{LOS}^2} = \displaystyle{\frac{\frac{1}{q_1^2}\frac{1}{q_2^2}}
                                          {\frac{\cos^2(i)}{q_1^2} + \frac{\sin^2(i)}{q_2^2}}
                                    }. 
\label{eqn:el}
\end{equation}
The cluster elongation index $q_{LOS}$ that is used elsewhere in the
paper (notation borrowed from \cite{betty05}) differs from $e_{LOS}$
in that $q_{LOS}$ is the ratio along the LOS to the photometric major
axis.

Thus, for oblate systems, when $q_1=1$, $q_2 = q > 1$, $q_p > 1$ and
$q_{LOS}=e_{LOS}$,
\begin{equation}
q_p^2 =\displaystyle{\frac{q^2}{q^2\cos^2(i) + \sin^2(i)}}
\end{equation}
Using this in Equation~\ref{eqn:el}, we get that for oblate systems,
\begin{equation}
q_{LOS}^2 =  \displaystyle{\frac{1 - q_p^2\cos^2(i)}{\sin^2(i)}}
\end{equation}

For prolate systems, $q_1=1$, $q_2 <1\Longrightarrow q_2 = 1/q$, $q_p < 1$ and $q_{LOS}=e_{LOS}/e_p$. So,
\begin{equation}
q_p^2 =\displaystyle{\cos^2(i) + q^2\sin^2(i)}
\end{equation}
Using this in Equation~\ref{eqn:el}, we get that for prolate systems,
\begin{equation}
q_{LOS}^2 =  \displaystyle{\frac{q_p^2 - \cos^2(i)}{q_p^4\sin^2(i)}}
\end{equation}

Since $q_p$ and $q_{LOS}$ are observables, $q_1$ and $q_2$ can be
calculated for the three different types of triaxial systems, as long
as we keep in mind that:
\begin{itemize}
\item for Triaxial-Type~I: $q_1 < 1$, $q_2 > 1$, $q_p < 1$ and
$q_{LOS} = e_{LOS}/q_p$.
\item for Triaxial-Type~II: $q_1 < 1$, $q_2 > 1$, $q_p > 1$ and
$q_{LOS} = e_{LOS}$.
\item for Triaxial-Type~III: $q_1 > 1$, $q_2 < 1$, $q_p > 1$ and
$q_{LOS} = e_{LOS}$.
\end{itemize}

We use Equations~\ref{eqn:ep} and \ref{eqn:el} to determine $q_1$ and
$q_2$ in a triaxial configuration. To do this, the values of $q_p$ and
$q_{LOS}$ will of course need to be supplied from observations.  The
value of $q_{LOS}$ that we consider is the centroid of the reported
error band in \cite{betty05}. 

However, for the different types of triaxial systems that we consider,
there are two other constraints involving $q_1$ and $q_2$ (in the form
of inequalities); such inequalities are typical of the triaxial type
of the cluster at hand and are listed above for Types~I, II and
III. Implementing these would imply two distinct ranges of
inclinations for each cluster. The overlap of these provides the final
constraints on the inclinations. Once this final range of inclinations
is obtained, Equations~\ref{eqn:ep} and \ref{eqn:el} are solved to
give $q_1$ and $q_2$, at the median of the recovered inclination
range.

\end{document}